  \documentstyle[12pt]{article}
  
\catcode`@=11
\def\secteqno{\@addtoreset{equation}{section}%
\def\theequation{\thesection.\arabic{equation}}}
\catcode`@=12
\secteqno
\newcommand{\be}{\begin{equation}}
\newcommand{\ee}{\end{equation}}
\newcommand{\bea}{\begin{eqnarray}}
\newcommand{\eea}{\end{eqnarray}}
\newcommand{\bref}[1]{(\ref{#1})}
\newcommand{\ep}{\epsilon} \newcommand{\vep}{\varepsilon}
\newcommand{\vf}{\varphi}
 
\newcommand{\f}{\phi}   \newcommand{\p}{\psi}
\newcommand{\A}{\alpha} 
\newcommand{\G}{\Gamma} \newcommand{\D}{\delta}

\newcommand{\F}{\Phi}           \newcommand{\V}{\Psi}

\def\pa{\partial}

\def\CD{{\cal D}}

\def\Kb{{\overline K}}
\def\psb{{\overline\psi}}
\def\Psb{{\overline\Psi}}

\newcommand{\nn}{\nonumber}

\begin{document}
          \hfill NIIG-DP-99-2

	  \hfill December, 1999

	  \hfill hep-th/9912262

\vskip 20mm 
\begin{center} 
{\bf \Large Exact Symmetries realized on the Renormalization Group Flow} 

\vskip 10mm
{\large Yuji\ Igarashi, Katsumi\ Itoh and Hiroto\ So$^a$}\par

\medskip
{\it 
Faculty of Education, Niigata University, Niigata 950-2181, Japan\\
$^a$ Department of Physics, Niigata University, Niigata 950-2181, Japan\\
}

\medskip
\date{\today}
\end{center}
\vskip 10mm
\begin{abstract}

We show that symmetries are preserved exactly along the (Wilsonian)
renormalization group flow, though the IR cutoff deforms concrete forms
of the transformations.  For a gauge theory the cutoff dependent
Ward-Takahashi identity is written as the master equation in the
antifield formalism: one may read off the renormalized BRS
transformation from the master equation.  The Maxwell theory is studied
explicitly to see how it works.  The renormalized BRS transformation
becomes non-local but keeps off-shell nilpotency.  Our formalism is
applicable for a generic global symmetry.  The master equation
considered for the chiral symmetry provides us with the continuum analog
of the Ginsparg-Wilson relation and the L{\" u}scher's symmetry.

\end{abstract}

\newpage
\setcounter{page}{1}
\setcounter{footnote}{0}
\parskip=7pt
\section{Introduction} 

The Wilsonian Renormalization Group (RG) is one of the most important
achievements in modern physics\cite{WilsonKogut}. In particular, the
exact RG equations[2-4] have proved to be powerful both in perturbative
and non-perturbative studies of field theories\footnote{See eg
\cite{Morris0}\cite{Aoki} for non-perturbative studies, [7-9] for
reviews of the recent development.}.  In a field theory, quantum
fluctuations at shorter distances are integrated out to give an
effective action for longer distances. For the well-defined integration,
one needs to introduce some regularization procedure, which may be in
conflict with symmetries in many important applications: for example,
the presence of gauge symmetry or chiral symmetry is far from trivial.
The incompatibility of symmetries and regularizations is a longstanding
problem in the RG approach.

There have been several attempts[10-13] recently to investigate this
problem based on a common recognition: a symmetry is broken at
intermediate steps of the RG iteration, and is recovered only after the
IR cutoff $k$ is removed.  The breaking of the symmetry is controlled by
the modified Ward-Takahashi (WT) identity, $\Sigma_k=0$\cite{Ellwanger}. 
In practical calculations, one has to finely tune parameters in an
effective action so that it satisfies the usual WT identity in the limit
of $k \rightarrow 0$.  This viewpoint, recovery of the symmetry by
``fine tuning", is due to Becchi\cite{Becchi} and extensively studied in
refs. \cite{Bonini}, \cite {DAttanasio} and \cite{Ellwanger}.

Recent development in understanding chiral symmetry on the lattice has
brought another important clue to our problem: L{\" u}scher found an
exact chiral symmetry on the lattice\cite{Luescher}, relying on the
Ginsparg-Wilson (GW) relation\cite{GW}. This provides us with the first
non-trivial example of having an exact symmetry even after the
regularization.  The L{\" u}scher's chiral symmetry takes quite
different form from that in the continuum limit.

Based on these observations, we shall give in this paper a general
method to define global symmetry along a RG flow.  It may be non-local
and cutoff dependent, yet exact symmetry even for $k \ne 0$.  We call
this ``renormalized symmetry''. Remarkably, our discussion applies to
gauge symmetry as well by considering its global counterpart, the BRS
symmetry.

We begin with a microscopic or UV action which is local and invariant
under a symmetry transformation. In order to construct the effective
action at low momentum, we consider the continuum analog of the
blockspin transformation. This formalism developed in\cite{Wetterich}
introduces macroscopic fields (average fields), in terms of which the
renormalized symmetry is realized. The important role of the
macroscopic fields is also suggested by the GW relation and the L{\"
u}scher's chiral symmetry.
Since the blockspin transformation is a gaussian integral, we obtain an
exact RG flow equation\cite{Wetterich} for the effective action of the
macroscopic fields. When expressed by the macroscopic fields and some
source fields, the WT identity $\Sigma_k=0$ takes the form of a master
equation, from which we shall find the exact symmetry transformation for 
$k \ne 0$.  We would like to emphasize here that our WT identity is for
the exact renormalized symmetry, not for the broken or modified symmetry.
This is the central issue of our formulation of renormalized
symmetry.  The flow equation for $\Sigma_k$ holds as a result of the
algebraic relation between the operator specifying the RG flow and that
appeared in the WT identity.

For gauge theories, the master equation is nothing but the one in the
antifield formalism of Batalin and Vilkovisky\cite{Batalin}.  In order
to see how the renormalized symmetry looks like, we give an effective
action and a renormalized BRS symmetry for the Maxwell theory.  As
another test of our method, we consider chiral symmetry, and show that
our master equation and associated renormalized symmetry are the
continuum analog of the GW relation and the L{\" u}scher's symmetry.  In
our derivation the GW relation is regarded as an exact WT identity for
the chiral symmetry.

\section{General formalism}

Let $\vf^A$ be a microscopic field\footnote{The index $A$ denotes kinds
of fields and other indices as a whole, except field momentum.}with the
Grassmann parity $\ep(\vf^A)=\ep_A$ and $S[\vf^A]$ a generic action.
The microscopic or UV action is assumed to be invariant $\D^{a} S[\vf]
=0$ under an infinitesimal global transformation with parameters
$\vep^{a}$, $\vf^A \to \vf^A + \D^{a}\vf^A\vep^{a} $, where
$\ep(\D^{a})= \ep(\vep^{a})$. The discussion to be given also applies to
gauge theory: the action $S[\vf]$ is a gauge fixed action and the
relevant global transformation is the BRS transformation.

To specify a blockspin transformation, we introduce a function
$f_{k}(p)$ with an IR cutoff $k$ in the Euclidean momentum space, and an
invertible matrix $[R_k(p)]_{AB}$ satisfying
$\ep([R_k(p)]_{AB})=0,~[R_k(p)]_{AB}=(-1)^{\ep_{A}\ep_{B}}[R_k(p)]_{BA}$. For
a boundary condition, we impose $f_{k}(p) \to 1,~ [R_k(p)]_{AB} \to
\infty$ as $k \to \infty$.  Possible choices of $f_{k}(p)$ and
$[R_k(p)]_{AB}$ were discussed in \cite{Wetterich}, but we do not need
to specify them here.  Let $K_{a}^A$ be sources for the variations
$\D^{a} \vf^A$: they will play an important role in our symmetry
consideration.  We may define an effective action for the macroscopic
fields $\F^A$ in the presence of the sources by
\bea
e^{-\G_{k}[\F,K]} &=& \int \CD \vf e^{-S_{k}[\vf,\F,K]},  \nn\\
S_{k}[\vf,\F,K] &=& S[\vf] 
+ \frac{1}{2}(\F - f_{k}\vf)^T_{-}~R_k~(\F - f_{k}\vf)_{+}
+ K_{a-}^{T}\D^{a}\vf_{+},
\label{gamma}  
\eea
where $\Phi_{\pm} \equiv \Phi(\pm p)$ and their multiplication implies the integration over momentum as well as the sum over the index $A$, eg,
\bea
\F^{T}_{-} R_k \F_{+} = {\rm Str}(R_k \F_{+} \F^{T}_{-}) 
\equiv \int_{p} \F^A(-p) [R_k(p)]_{AB} \F^B(p),~~~~
\int_{p} \equiv \int d^D p.
\label{matrix}
\eea
The supertrace, Str, denotes a sum over momenta and indices. Note that
$f_{k}[R_k]_{AB}\F^B$, a linear term of the macroscopic fields, acts
as a source term for $\vf^A$ in the path integral.

Since only the gaussian term depends on the cutoff $k$, one obtains the
exact RG flow equation\cite{Wetterich} for the macroscopic action
$\G_{k}[\F,~K]$ :
\bea
\pa_{k}e^{-\G_{k}[\F,K]}&=& - \left[X + \frac{1}{2} {\rm Str}(R_k^{-1}
\pa_{k}R_k {\bf 1}) + {\rm Str}(\pa_{k}(\ln f_{k}){\bf 1})\right] e^{-\G_{k}[\F,K]}, \nn\\
X &\equiv&   -\frac{1}{2} \frac{\pa^l}{\pa
\F^{T}_{-}}(\pa_{k}R_k^{-1})\frac{\pa^r}{\pa\F_{+}}+
\pa_{k}(\ln f_{k})\left[\frac{\pa^l}{\pa\F^T_{-}}R_k^{-1}\frac{\pa^r}{\pa\F_{+}}+
\F^{T}_{-}\frac{\pa^l}{\pa\F_{+}}\right]. 
\label{flow}
\eea

We consider now the symmetry property of the macroscopic action.
Invariance of the microscopic action under the global transformation can
be expressed as
\bea
\int \CD \vf e^{-S_{k}[\vf +\D^{a}\vf\vep^{a},\F,K]}
~-~\int \CD \vf e^{-S_{k}[\vf,\F,K]}=0.
\label{WT1}
\eea
Assumed here is the translational invariance of the path integral
measure, ie, the absence of anomalies.  For each independent parameter
$\vep^{a}$, the WT identity reads
\bea
\Sigma_{ka}[\F,K] \equiv -e^{\G_{k}[\F,K]} \Delta_{a} e^{-\G_{k}[\F,K]}
= - \langle K_{b-}^{T} \D_{a} \D^{b}\vf_+ \rangle_k,
\label{Sigma1}
\eea
where the expectation value is taken with respect to the action $S_k$
and the operator $\Delta_a$ is defined by
\bea
\Delta_{a} \equiv {\rm Str} \left(f_{k} \frac{\pa^{l}}{\pa \F_{-} }
\left( \frac{\pa^{l}}{\pa K^{a}_{+}}\right)^T \right).
\label{lap}
\eea
One obtains 
\bea 
\Sigma_{ka} = {\rm Str} \left( f_{k}
\frac{\pa^{l}\G_{k}} {\pa \F_{-} }
\left(\frac{\pa^{l} \G_{k}}{\pa K^{a}_{+}} \right)^T\right)
+\Delta_{a}\G_k. 
\label{Sigma2} 
\eea 
This takes the form of a master equation in the space of
$(\F^{A},~K^{a}_{A})$. As will be seen presently, for the BRS symmetry
the source $K_{A}(p)/f_{k}(p)$ can be identified with the antifield of
the macroscopic field $\F^{A}$, and \bref{Sigma1} becomes the quantum
master equation.

In oder to obtain the flow equation for $\Sigma_{ka}$ in our
formulation, we notice that there is an algebraic relation between
operator $X$ in \bref{flow} and the operator $\Delta_a$:
\bea 
[\Delta_{a}, ~X] = (\pa_k \Delta_{a})
\label{algebra} 
\eea 
on any Grassmann even quantity.
This leads to the flow equation\cite{Ellwanger}
\bea \pa_{k} \Sigma_{ka} = (e^{\G_{k}}X
e^{-\G_{k}})\Sigma_{ka} -
e^{\G_{k}}X~(e^{-\G_{k}}\Sigma_{ka}). 
\label{Ell} 
\eea 
It is easily seen that the r.h.s consists of the functional
derivatives of $\Sigma_{ka}$.

The above equations \bref{Sigma1} $\sim$ \bref{Ell} hold quite
generally.  They also provide us with the transformation for the
renormalized symmetry.  In the following two sections we consider the
BRS and global symmetry separately.

\section{Renormalized BRS symmetry}

\subsection{The master equation}

For the BRS symmetry, the source $K_{A}(p)/f_{k}(p)$ can be identified
with the antifield $\F_A^{*}$ for the macroscopic field $\F^A$.  Then,
the operator $\Delta$ in \bref{lap} and a bracket defined by
\bea 
(F,~G)\equiv \int_{p} \left[\frac{\pa^{r}
F}{\pa\F^{A}(-p) }\frac{\pa^{l} G}{\pa \F_A^{*}(p)} - \frac{\pa^{r}
F}{\pa \F_A^{*}(-p)}\frac{\pa^{l} G}{\pa \F^{A}(p)}\right],  
\label{bra}
\eea 
are exactly those in the antifield formalism of
Batalin-Vilkovisky\cite{Batalin}. Since the r.h.s of \bref{Sigma1}
vanishes because of the nilpotency $\D^2$=0, one obtains the condition,
\bea
\Sigma_{k}[\F,\F^{*}]= \frac{1}{2}
(\G_{k}[\F,\F^{*}],~\G_{k}[\F,\F^{*}]) + \Delta \G_{k}[\F,\F^{*}]=0,
\label{SigmaBRS}
\eea
which is nothing but the quantum master equation.  It is an algebraic
equation which holds for any $\F$ and $\F^{*}$.  The WT flow equation
\bref{Ell} tells us then that once it is satisfied at some cutoff
$k=k_0$ it persists along the RG flow.  This clearly demonstrates
the presence of a cutoff dependent BRS symmetry, a renormalized BRS
symmetry, in the macroscopic action. If the second term in the master
equation vanishes, we may define the renormalized BRS transformation on
$\Phi$ and $\Phi^*$ by
\bea
{\delta_{r}} \F^{A}= (\F^{A}, \G_k[\F,\F^{*}]),~~~
{\delta_{r}} \F_{A}^*= (\F_{A}^*, \G_k[\F,\F^{*}]).
\label{BRS2}
\eea

The cutoff dependent BRS transformation appeared earlier in a different
context\cite{Becchi}.  The author took the viewpoint to finely tune the
effective action for $k \ne 0$ with gauge non-invariant terms so that it
satisfies the usual WT identity in $k \rightarrow 0$ limit.  A series of
papers followed to confirm this point of view perturbatively for various
models\cite{Bonini}.  The ``modified Slavnov-Taylor identity'' and its
flow equation are elegantly summarized in \cite{Ellwanger}.  However the
presence of the exact BRS symmetry had not been understood.

Here we have seen that the transformation may be defined with the master
equation in the antifield formalism, and the WT identity $\Sigma_{k}=0$
is not a broken but exact identity.  In the next subsection we shall
give a simple model of the renormalized BRS symmetry for the Maxwell
theory, where the above stated properties can be confirmed explicitly.

\subsection{Abelian gauge symmetry}

Let us consider the gauge-fixed Maxwell action in D=4 Minkowski space,
\begin{eqnarray}
S_0[\varphi, \vf^{*}] = \int \left[
-\frac{1}{4}F^2 + B(\partial \cdot A + \frac{\alpha}{2} B) 
+ i \partial^{\mu}{\bar c} \partial_{\mu} c + \vf^{*T} \delta\varphi
\right],
\label{S0}
\end{eqnarray}
where
\begin{eqnarray}
\varphi \equiv \left(
\begin{array}{c}
 A_{\mu} \\ c\\ {\bar c}\\ B
\end{array}
\right),~~~
\delta\varphi = \left(
\begin{array}{c}
 \partial_{\mu}c \\ 0\\ iB\\ 0
\end{array}
\right),~~~
\vf^{*} \equiv \left(
\begin{array}{c}
 A_{\mu}^{*} \\ c^{*}\\ {\bar c}^{*}\\ B^{*}
\end{array}
\right).
\label{micro}
\end{eqnarray}
The microscopic action $S_{0}$ satisfies the (classical) master
equation, $(S_{0}, S_{0})=0$, for the antibracket defined in terms of
$\vf$ and $\vf^{*}$: the $\varphi^*$ is the set of the antifields at the
microscopic level.\footnote{Note that the BRS transformation in
\bref{S0} is defined by the right derivative: $\delta \vf^{A} =
(\vf^A,~S_{0})$.}  The macroscopic fields,
\begin{eqnarray}
\Phi \equiv \left(
\begin{array}{c}
 {\cal A}_{\mu} \\ {\cal C}\\ {\bar {\cal C}}\\ {\cal B}
\end{array}
\right),
\label{macro}
\end{eqnarray}
have an effective action defined in the relation, 
\begin{eqnarray}
e^{i \Gamma_k[\Phi,\vf^{*}]} = \int {\cal D} \varphi e^{i S_k[\varphi,\Phi,\vf^{*}]},
\label{defGammak}
\end{eqnarray}
where
\begin{eqnarray}
S_k[\varphi,\Phi,\varphi^*] \equiv S_0[\varphi,\vf^{*}]+ 
\frac{1}{2} (\Phi-f_k\varphi)_{-}^T~R_k~(\Phi-f_k\varphi)_{+}
\label{Sk}
\end{eqnarray}
with
\begin{eqnarray}
R_k(p) \equiv M_k^{2}(p) \left(
\begin{array}{cccc}
 g^{\mu \nu} & & &  \\  & & i& \\  & -i& & \\ & & & 1/{\mu_k^2(p)}
\end{array}
\right).
\label{kernel}
\end{eqnarray}
We have chosen the blockspin kernel, the gaussian term, like a mass term
: both $M_k(p)$ and $\mu_k(p)$ have the dimension of mass.

All the terms are bilinear so that we may obtain the macroscopic action
explicitly,
\begin{eqnarray}
\Gamma_k[\Phi,\vf^{*}] = 
\frac{1}{2}(\Phi_{-}^T,K_{-}^T[\vf^{*}])
\left(
\begin{array}{cc}
R_k-f_k^2 R_kD^{-1}R_k& f_k R_k D^{-1}\\f_k D^{-1} R_k& -D^{-1} 
\end{array}
\right)
\left(
\begin{array}{c}
\Phi_{+}\\ K_{+}[\vf^{*}]
\end{array}
\right),
\label{Gammak3}
\end{eqnarray}
where $D(p)$ is the matrix defined in the relation,
$S_0[\varphi,\varphi^*]=\frac{1}{2}\varphi_{-}^T(D-f_k^2R_k)\varphi_{+}+\varphi^{*T}_{-}\delta
\varphi_{+}$, and $K_{\pm}[\vf^{*}]$ are the compact notations for the
following vectors,
\begin{eqnarray*}
K_{\pm}[\vf^{*}] \equiv \left(
\begin{array}{c}
0\\ -i~p \cdot A^{*}(\pm p)\\ 0\\ i {\bar c}^{*}(\pm p)
\end{array}
\right).
\end{eqnarray*}
%


Since $\Delta \Gamma_{k}=0$, one obtains the renormalized BRS
transformation for $\Phi$ as
\begin{eqnarray}
{\delta_r} {\cal A}_{\mu}(p) 
&=& f_k \frac{\partial^l \Gamma_k}{\partial A^{*}(-p)}
= i p_{\mu} a(p) (f_k M_k)^2 {\cal C}(p),
\nonumber\\
{\delta_r} {\bar {\cal C}}(p)
&=& f_k \frac{\partial^l \Gamma_k}{\partial {\bar c}^{*}(-p)}
= (f_k M_k)^2 b(p) \left[ p \cdot {\cal A}(p) -i \left(\frac{f_k M_k}{\mu_k}\right)^2 {\cal B}(p) - f_k {\bar c}^{*}(p) \right],
\nonumber\\
\delta_r {\cal C}(p)&=&\delta_r {\cal B}(p)=0, 
\label{rBRS}
\end{eqnarray}
where
\begin{eqnarray*}
a(p) \equiv \frac{1}{p^2-(f_k M_k)^2},~~~
b(p) \equiv \frac{1}{p^2-(f_k M_k)^2[\alpha+(f_k M_k / \mu_k)^2]}.
\end{eqnarray*}
In spite of the non-locality and the operator mixing, the renormalized
BRS transformation is nilpotent on ${\bar {\cal C}}(p)$, which may be
easily confirmed once we take account of the transformation of the
antifield ${\bar c}^*(p)$,
\begin{eqnarray}
{\delta_r} {\bar c}^{*}(p)   
&=& - f_k \frac{\partial^l \Gamma_k}{\partial {\bar {\cal C}} (-p)}
= i f_k M_k^2 a(p) p^2 {\cal C}(p).
\label{rBRS2}
\end{eqnarray}
With similar calculations, one can obtain renormalized BRS
transformations of other antifields and explicitly observe their
(off-shell) nilpotency.

The above mentioned complication of $\delta_r$ is partly caused by the
bilinear term in the antifields appeared in $\Gamma_k$.  A canonical
transformation may be used to make the action linear in antifields,
which will not be discussed further.

\section{Renormalized global symmetry}

We now discuss other global symmetry.  In this case, the r.h.s. of
\bref{Sigma1} does not vanish in general. Therefore, to obtain the WT
identity we need to set $K=0$ after taking the functional derivatives in
\bref{Sigma1}:
\bea \Sigma_{ka}[\F] =
\left[{\rm Str} \left( f_{k}
\frac{\pa^{l}\G_{k}} {\pa \F_{-} }
\left( \frac{\pa^{l} \G_{k}}{\pa K^{a}_{+}} \right)^T \right)
+\Delta_{a}\G_k \right]_{K=0}
=0. 
\label{Sigma3} 
\eea 
This is an algebraic relation to hold for any $\F$. Unlike the BRS
symmetry, we have no natural bracket structure\footnote{For a Grassmann
odd symmetry such as supersymmetry, however, we may define a bracket
in the space of $(\F^{A},~K^{a}_{A})$.}.  Yet, since the operator $X$
contains no functional derivatives with respect to $K$, the WT flow
equation \bref{Ell} is unchanged: it persists again along the RG
flow. Thus, the quantum master equation \bref{Sigma3} ensures the
presence of renormalized global symmetry. In the absence of the
$\Delta_a \Gamma_k$ term, the transformation for the renormalized
symmetry is given by
\bea
{\delta_r} \F^{A}(p)= f_{k}(p) \left[\frac{\pa^{l} \G_{k}}{\pa
K^{a}_{A}(-p)}\right]_{K=0}.  
\label{globaltra} 
\eea

We now apply our formalism to the chiral symmetry. Let
$(\p,~\psb),~(\V,~\Psb)$ be microscopic and macroscopic fermion fields,
respectively. We introduce the sources $(\Kb,~K)$ for the variations
$(\D \p = i \gamma_{5}\p,~\D \psb= i \psb \gamma_{5})$. The macroscopic
action is given by
\bea
e^{-\G_{k}[\V,\Psb,K,\Kb]} &=& \int \CD \p \CD \psb 
e^{-S_{k}[\p,\psb,\V,\Psb,K,\Kb]},  \nn
\eea
where
\bea
&{ }&S_{k}[\p,\psb,\V,\Psb,K,\Kb]
\nn\\
&=& S[\p,\psb]
+ (\Psb -f_{k}\psb )_{-}\A_{k}
(\V -f_{k}\p )_{+}
-\psb_{+}  i \gamma_{5} K_{-} + \Kb_{-} i \gamma_{5} \p_{+}. 
\label{gammachiral}  
\eea
$\A_{k}$ is a function of $k$ and $p$.
The gaussian contains linear terms in $\p$ and $\psb$:
\bea
-\A_{k}f_{k}\left[\left(\Psb - (\A_{k}f_{k})^{-1}\Kb i \gamma_{5}\right)_{-}
\p_{+} + \psb_{-} \left(\V + (\A_{k}f_{k})^{-1}i \gamma_{5}
K\right)_{+}\right].
\label{linear}
\eea
Generically these are the only terms which act effectively as sources
for $(\p,~\psb)$ in the path integral. If we assume the macroscopic
action to be bilinear in the macroscopic fermions, it takes the form,
\bea
\G_{k}[\V,\Psb,K,\Kb]&=& \bigl(\Psb -
(\A_{k}f_{k})^{-1}\Kb i \gamma_{5}\bigr)_{-} (D - \A_{k})\bigl(\V  + (\A_{k}f_{k})^{-1}i \gamma_{5}
K\bigr)_{+} \nn\\
&{}&~~~~ + \Psb_{-} \A_{k} \V_{+},
\label{bilinear}
\eea
where $D$ denotes the Dirac operator for the macroscopic fields, defined as
the coefficient of $\Psb_{-}~\V_{+}$. Then the master equation
\bref{Sigma3} gives
\bea
\Sigma_{k}[\V,~\Psb]= i \Psb_{-}\left[D \gamma_{5}
(1- \A_{k}^{-1} D) +  (1- \A_{k}^{-1} D)\gamma_{5} D  \right]\V_{+} 
=0
\label{Sigmachiral}
\eea
where we have used ${\rm tr}\{\gamma_{5},~D\}=0$, which is legitimate in
the absence of chiral anomalies. One obtains in this way the continuum
analog of the GW relation:
\bea
\{\gamma_{5},~D\} = 2 \A_{k}^{-1} D~\gamma_{5}~D.
\label{GW}
\eea
  
Since the second term in \bref{Sigma3} vanishes owing to ${\rm
tr}\{\gamma_{5},~D\}=0$, the chiral transformation on the macroscopic
fields is readily given by
\bea
{\delta_r} \Psi &=& f_k \frac{\partial^l \Gamma_k}{\partial {\overline K}}
= i \gamma_5(1- \A_{k}^{-1} D) \Psi, \nn\\
{\delta_r} {\overline \Psi} &=& f_k \frac{\partial^l \Gamma_k}{\partial K}
= i {\overline \Psi}(1- \A_{k}^{-1} D)\gamma_5,
\eea
which is nothing but the L{\" u}scher's symmetry transformation.  For
the chiral symmetry, therefore, the master equation $\Sigma_{k}=0$ is
identified with the GW relation. The flow equation \bref{Ell} tells us
that it persists along the RG flow. The L{\" u}scher's symmetry turns
out to be the renormalized symmetry realized on the flow.  It is
probably worth pointing out that the variants in L{\" u}scher's
symmetry\cite{Luescher} are naturally understood in our formulation: an
arbitrary vector perpendicular to ${\partial^r \Gamma_k}/{\partial
\Phi^A}$ may be added to the transformation since it does not change the
condition $\Sigma_k=0$.

So far we have discussed for $\D_{a} S[\vf]=0$. Before closing this
section, let us consider briefly the microscopic action with some
non-invariant terms, $\D_{a} S[\vf]\neq 0$. The presence of $\D_{a}
S[\vf]$ gives a new contribution in \bref{Sigma1}. Now $\Sigma_{ka}[\F]$
does not vanish even after taking $K=0$. It should be remarked however
that the non-vanishing term defined by
\bea
\sigma_{ka}[\F] = - \langle \D_{a} S[\vf] \rangle_k 
\label{Sigmabreak}
\eea
still satisfies the flow equation
\bea
\pa_{k} \sigma_{ka} = (e^{\G_{k}}X~ e^{-\G_{k}})\sigma_{ka}
-e^{\G_{k}}X~(e^{-\G_{k}}\sigma_{ka}). 
\label{Ellbreak2}
\eea
This equation gives us some important information on the RG flow of the
couplings for the non-invariant terms.  It is straightforward to extend
the eqs.\bref{Sigmabreak} and \bref{Ellbreak2} to the case of the BRS
symmetry.

\section{Summary}

We have shown that a symmetry, not compatible with a given
regularization, may survive exactly along the RG flow.  The concrete
realization of the symmetry reflects deformation due to the
regularization.\footnote{The realization of symmetries in such a form
has been considered for the lattice gauge theory as one of the
properties of the perfect action\cite{Hasenfratz}.}  Naturally it
reduces to the usual form in the $k \to \infty$ limit.  In this letter
we have presented a general formalism based on the ``average action'', a
continuum cousin of the blockspin transformation.  The WT identity for
the renormalized symmetry takes the form of the master equation, from
which we may read off the associated transformation on the macroscopic
fields.

The Maxwell theory was found to be a simple yet instructive example to
understand the renormalized BRS transformation.  As a result of the
blockspin transformation it became non-local but still kept the
off-shell nilpotency, as it should from our general argument.  For the
chiral symmetry in a continuum theory, we have identified the GW
relation with the WT identity $\Sigma_{k}=0$.  Our formalism naturally
leads us to identify the renormalized chiral symmetry with the L{\"
u}scher's symmetry. This is regarded as another non-trivial example of
the renormalized symmetries.

We are grateful to M. Kato and H. Nakano for discussions on related
subjects.


\vspace{0.5cm}
\begin{thebibliography}{99}
\bibliographystyle{unsrt} 

%
\setlength{\itemsep}{0.0in}

\bibitem{WilsonKogut} K. G. Wilson and J. Kogut,
Phys. Rep. {\bf C12} (1974) 75.

\bibitem{WegnerHoughton} F. J. Wegner and A. Houghton,
Phys. Rev. {\bf A8} (1973) 401.

\bibitem{Polchinski} J. Polchinski, 
Nucl. Phys. {\bf B231} (1984) 269.

\bibitem{Wetterich} C. Wetterich,
Nucl. Phys.{\bf B352} (1991) 529; Z. Phys. {\bf C60} (1993) 461.

\bibitem{Morris0} T. R. Morris, Int. J. Mod. Phys. {\bf A9} (1994) 2441.

\bibitem{Aoki} K-I. Aoki, K. Morikawa, W. Souma, J-I Sumi and H. Terao, 
Prog. Theor. Phys. {\bf 95} (1996) 409.

\bibitem{Morris} Tim R. Morris, 
{\it New Developments in the Continuous Renormalization Group}, 
Invited key talk at NATO Advanced Research Workshop on Theoretical Physics: 
 New Developments in Quantum Field Theory, Zakopane, Poland, 
14-20 Jun 1997, hep-th/9709100.

\bibitem{YKIS97} 
Proc. of the 8th Yukawa International Seminar, {\it Non-perturbative
		QCD},
eds. K-I Aoki, O. Miyamura and T. Suzuki, 
Prog. Theor. Phys. Suppl. {\bf 131} (1998).

\bibitem{Vian} F. Vian, hep-th/9905142.

\bibitem{Becchi} C. Becchi,
{On the construction of renormalized quantum field theory using
		renormalization group techiniques}, in {\it Elementary
		particles, Field theory and Statistical mechanics}, eds. M. Bonini, G. Marchesini and E. Onofri, Parma University 1993.

\bibitem{Bonini} M. Bonini, M. D'Attanasio and G. Marchesini,
		Nucl. Phys. {\bf B418} (1994) 81; {\it ibid} {\bf B421}
		(1994) 429; {\it ibid} {\bf B437} (1995) 163;
Phys. Lett. {\bf B346} (1995) 87; M. Bonini and G. Marchesini,
		Phys. Lett. {\bf B389} (1996) 566.

\bibitem{DAttanasio} M. D'Attanasio and T. R. Morris, 
Phys. Lett.  {\bf B378} (1996) 213.

\bibitem{ReuterWetterich} M. Reuter and C. Wetterich,
Nucl. Phys. {\bf B 417} (1994) 181; {\it ibid} {\bf B 427} (1994) 291;
F. Freire and C. Wetterich, Phys. Lett. {\bf B380} (1996) 337.

\bibitem{Ellwanger} U. Ellwanger,
Phys. Lett. {\bf B335} (1994) 364.

\bibitem{Luescher} M. L{\" u}scher,
Phys. Lett. {\bf B428} (1998) 342; Nucl. Phys. {\bf B549} (1999) 295.

\bibitem{GW} P. Ginsparg and K. Wilson, 
Phys. Rev. {\bf D25} (1982) 2649.

\bibitem{Batalin} I. A. Batalin and G. A. Vilkovisky,
Phys. Lett. {102B} (1981) 27.

\bibitem{Hasenfratz} P. Hasenfratz,
{\it The theoretical background and properties of perfect actions},
in Proc. of the workshop, {\it Non-perturbative quantum field physics},
		Peniscola 1997, hep-lat/9803027.

\end{thebibliography}
\end{document}